\documentclass[a4paper,12pt]{article}
\usepackage[cp1251]{inputenc}
\usepackage [russian] {babel}
\usepackage [dvips] {graphicx}
\usepackage[]{caption2}
\usepackage{floatflt}
\usepackage{bm}
\usepackage{amsmath}
\usepackage[usenames]{color}
\usepackage {amssymb} 

\title {\Large\textbf{rn}}

\date{}

\title{Kinetic phase diagram  for a binary system near the
transition to diffusionless solidification}
\begin{document}
\maketitle
\renewcommand{\figurename}{Fig}

\begin{center}
\author{G.L. Buchbinder\footnote{Email:
glb@omsu.ru}} \end{center}
 \small{  Physics Department, Omsk State
University,  Pr. Mira 55a, Omsk, 644077, Russian Federation}

\begin{abstract} The rapid solidification of a binary mixture in the
region of the interface velocities $V$ close to the diffusion speed
in the bulk of the liquid phase $V_D$ is considered within the
framework of the local nonequilibrium approach.  In this high-speed
region the derivation of the analytical expression for the response
function "temperature-velocity" representing kinetic phase diagram
is given without using the concept of the equilibrium phase diagram.
  The modes of movement of the
interface both without and with the drag effect are analyzed. It is
shown that the drag effect can be accompanied by a local  interface
temperature maximum at $V = V_D$.
\end{abstract}

\section{ Introduction}
 \label{int}
  Quantitative modeling of the process of nonequilibrium solidification of metal
  melts has been attracting attention for the past few decades and is an object
  of scientific interest at the present time \cite{GJ19,HGH07}.
   This interest is primarily due to the fact that nonequilibrium solidification provides
   the potential opportunity for obtaining metastable materials and, in particular, supersaturated solid solutions \cite{HGH07}.

At low speeds of the solidification front the composition and
properties of the final phase can be predicted based on
representations about local equilibrium of the interface and usage
of the equilibrium phase diagram.  However, at sufficiently high the
front speeds deviations from the local equilibrium at the interface
can become significant.   Rapid solidification experiments
\cite{A95, BT95,KA95,KS00, EC92} show that on a fast-moving the
phase boundary the processes of segregation are suppressed, the
solute distribution
 is no longer determined in accordance with the equilibrium phase
 diagram and the solute trapping  by the solid phase takes place.

The effect of complete solute trapping arises when the
solidification speed $V$ reaches the value of the solute diffusion
speed in the bulk of the liquid phase $V_D$  \cite{S95}.  In this
case, the solute does not have time to diffuse into the bulk of the
liquid and is trapped by the solid phase at a concentration equal to
its initial concentration in the melt.  The new phase is then formed
under conditions of diffusionless solidification with the partition
coefficient k = 1. A number of theoretical models proposed earlier
for describing solute trapping (see, for example, review
\cite{GJ19}) actually assume that $V/V_D \ll 1$ and predict that
complete solute trapping with $k = 1$ is only possible
asymptotically for $V \rightarrow \infty$, that corresponds to an
infinitely large value of $V_D$.  However available experimental
data \cite{BT95,KA95,KS00,EC92} clearly demonstrate that the
transition to complete solute trapping giving rise  to diffusionless
solidification occurs at essentially finite values of $V$.
 Theoretically, the same conclusion follows from the locally nonequilibrium model (LNM) developed in the works  \cite{S95,S96,S97,GS97,
G02,GD04,GH06,G07,S12,S15,S17}.  According to LNM, the
solidification speeds observed in modern experiments can reach or
even exceed the diffusion speed $V_D$, and as a result, the
deviation from local equilibrium can be significant not only at the
interface but in the bulk of the liquid phase as well.   The
so-called local nonequilibrium approach has been successfully
applied to numerical simulation of experimental data over a wide
range of solidification rates and is consistent with MD modeling of
rapid solidification of a number of binary systems
\cite{WL07,LZ10,LZ11,WL11,WL12, YH11}.

The modeling of the solidification process depends to a large extent
on the   boundary conditions realized at the phase interface and, in
 particular,  the relationship between temperature, the interface velocity
 and  the solute  concentration ( so-called the response  function  “temperature-velocity”).
The response function together with the nonequilibrium partition
coefficient $k(V)$ allows one to find the kinetic phase diagram
generalizing the equilibrium phase diagram  to    nonequilibrium
solidification process.

Within the frame of  LNM derivation an analytical expression for the
response function
 in the case of a dilute melt, sharp
interface and linear approximation for equilibrium solidus and
liquidus has been  considered in \cite{G02,S15}. In particular, it
has been assumed that at the temperature at which  the interface
moves, the solid and liquid phases can be in equilibrium with each
other with the solute concentrations determined by the equilibrium
solidus and liquidus, i.e. determined on the basis of the
equilibrium phase diagram \cite{BC86}. This approximation may be
applicable if undercooling at the solidification front is relatively
not large. However at large undercooling achievable in modern
experiments it may turn out that moving at high speed the interface
has a sufficiently low temperature at which an alloy in an
equilibrium state in the liquid phase does not exist and usage of
the above assumption is not correct. For example, during the
solidification of Si-9\, at.\% As alloy, often used as a test
system, the interface temperature can reach a value of the order $T
= 1373$K \cite{KS00}, below which a dilute alloy exists in
equilibrium  only in the form of a solid solution \cite{OA85}.

In this paper within the local nonequilibrium approach we consider
the derivation of the "temperature-velocity" response function
without using the concept of equilibrium phase diagram, i.e. without
using equilibrium properties systems, and without using the
assumption  about phase equilibrium at the temperature of the
fast-moving interface. The problem is considered in the region of
speeds $V$ close to the critical speed $V_D$ at which
 a transition from diffusion-controlled growth to
diffusionless solidification with $k = 1$ is occurred . The idea of
considering the solidification process in this speed region has been
previously proposed to find the  boundary interface conditions and
the nonequilibrium partition coefficient for the interface moving at
a constant temperature \cite{BG20}. Here, the approach developed in
\cite{BG20} is applied to take into account temperature changes at
the interface and to obtain the response function for large
supercooling in the case of a diluted melt. The response function
(nonequilibrium liquidus) is then used to analyze the temperature
behavior of the interface near the transition to diffusionless
solidification and compare the modes of movement of the interface
both with and without taking into account the drag effect.

\section{The  chemical potentials}
 Consider a solidifying binary alloy consisting of two
component $A$ (solvent) and $B$ (solute).  Let the molar fractions
of the solute on the liquid and the solid sides of the interface,
respectively, are $C_L$ and $C_S$. The liquid and solid phases are
separated by a planar interface moving steady-state with the
velocity of $V$. When modeling  the response function   one usually
proceeds from the kinetic equation  relating the velocity $V$ to the
thermodynamic driving force of phase transformation $\Delta G$
\begin{equation}\label{1}
    V = V_0(1 - e^{\Delta G/RT})\,,
\end{equation}
where $T$ is the interface temperature, $R$ is the gas constant and
$V_0$ is the upper limit of the solidification rate \cite{T62}. In
the model of solidification without solute drag effect  $\Delta G$
is represented as (see, for example, \cite{AK88})
\begin{equation}\label{2}
    \Delta G = \Delta G_{DF}  =  C_S\Delta\mu_B + (1 - C_S)\Delta\mu_A\,,
\end{equation}
where $\Delta G_{DF}$ is the change of the Gibbs free energy when
one mole of a substance solidifies in the composition  $C_S$,
$\Delta\mu_i = \mu_i^S - \mu_i^L$ and $\mu_i^{LS}$ are chemical
potentials of the component $i$ ($i = A, B$) at the liquid ($L$) and
the solid ($S$) sides of the interface. In the model with solute
drag $\Delta G$ is given by
\begin{equation}\label{3}
    \Delta G = \Delta G_C  =  \Delta G_{DF} - \Delta G_{D} = C_L\Delta\mu_B + (1 - C_L)\Delta\mu_A\,,
\end{equation}
where $\Delta G_C$ is the crystallization free energy and $\Delta
G_D = (C_L - C_S)(\Delta\mu_A - \Delta\mu_B)$ is   a part of the
free energy spent on changing the composition of the  solid phase
with $C_L$ on $C_S$ when $A$ and $B$ atoms diffuse through the
interface.

Using (\ref{2}) (or (\ref{3}))  the Eq. (\ref{1}) can be written  as
\begin{equation}\label{7}
   C_{S(L))}\Delta \mu_B + (1 - C_{S(L)})\Delta\mu_A = RT\ln (1 - V/V_0)\,.
\end{equation}
It should be noted that the driving force of the phase
transformation  $\Delta G$ at given changes $\Delta\mu_A$,
$\Delta\mu_B$ is uniquely defined by Eq. (\ref{2}) (or (\ref{3}) if
solute drag  is taken into account) The nature of the nonequilibrium
state in the bulk of phases is manifested  only through the chemical
potentials $\mu_i^{LS}$. If the medium is in a local equilibrium
state $\mu_i^{LS}$ are functions of the concentration $C$ and
temperature $T$. However, in the case of rapid motion of the
interface, the local equilibrium in the diffusion field of the
liquid phase may not have time to establish.  In this case  the
local nonequilibrium approximation is more appropriate. In a local
nonequilibrium state the entropy of the system $S$  depends on not
only  the classical variables $C$ and $T$ but also the diffusion
current $J$, which, like the concentration and temperature, is
considered as an independent variable, i.e., $S = S(C, T, J)$
\cite{JC96}. It follows that in a local nonequilibrium state the
chemical potential $\mu = - T\partial S/\partial C$ is a function of
the same variables, that is $\mu^{LS}_i = \mu^{LS}_i (C,T,J)$
\cite{BG20}.

Now let the interface move stationary at a speed $V$ close to the
velocity $V_D$ for which the diffusionless solidification  takes
place with  $J_L = 0$ and the  solute concentration  in both phases
equals to the initial concentration in the melt $c_0$.  Taking into
account the above and neglecting diffusion in the solid phase, one
can write down for chemical potentials in the region
 $V \sim V_D$
  the following expansion \cite{BG20}
\begin{eqnarray}
 \mu^L_i (C_L, T, J_L)&=&\mu^L_{leq,\, i}(C_L, T) + \alpha_i\frac{RT}{\rho V_D}J_L + \cdots \label{8} \\
  \mu^S_i (C_S, T, J_S)&=&\mu^S_{leq,\, i}(C_S, T)\,,  \label{9}
\end{eqnarray}
where  $\rho$  is  the mass density of the medium \cite{BG20}. The
terms in (\ref{8}) and (\ref{9})  independent on $J_L$  represent
the local equilibrium part of the chemical potential
$\mu^{LS}_{leq,\,i}$. As it follows from  Eq. (\ref{8}) for
$V_D\rightarrow \infty$, $\mu_i^L \rightarrow \mu_{leq,\, i}^L$. The
coefficient at  $J_L$ has been  chosen such that $\alpha_i$ is a
dimensionless parameter of the order of one  and its sign it can be
defined for the following reasons. According to a well-known
equality, for the entropy of the liquid phase one can write down
using Eq. (\ref{8})
\begin{equation*}
    S = - \sum\limits_{i}C_i\Bigg (\frac{\partial\mu_i^L}{\partial T}\Bigg )_{p,J} =
    S_{leq} - \frac{RJ_L}{\rho V_D}\sum\limits_i\alpha_iC_i\,,
\end{equation*}
where $p$ is the pressure. Since in the local equilibrium state the
entropy of the system $S_{leq}$ is greater than in the local
nonequilibrium one\footnote{An isolated thermodynamic system,
initially being
    in the local nonequilibrium state with entropy $S$, in the process of
    relaxation to an equilibrium state passes through an intermediate
the local equilibrium state. Since its entropy can only increase
during the irreversible process, $S < S_{leq}$ (see also discussion
in \cite{SU97}) }, then $\Delta S = S-S_{leq} < 0$. The latter
inequality is automatically executed when $\alpha_i > 0$ ($J_L >
0$).

For states close to diffusionless solidification the local
equilibrium part of the chemical potential can be expanded into a
series in powers of $C - c_0$ and $T - T^*$, where $T^*$ is the
temperature of the interface moving with a speed  $V_D$. Then,
restricting ourselves to the linear approximation, we write Eqs.
(\ref{8}) and (\ref{9}) in the following form
\begin{eqnarray}\label{10}
&& \mu_i^L(C_L, T, J_L) = {\displaystyle \mu_i^{*L}  +
\frac{\partial \mu_i^{*L}}{\partial c_0}(C_{L} - c_0)} \nonumber\\
&&\phantom{aaaaaaaaaa} +{\displaystyle\frac{\partial
\mu_i^{*L}}{\partial T^*}(T - T^*)} + {\displaystyle
\alpha_i\frac{RT}{\rho V_D} J_L}
\end{eqnarray}
\begin{equation}\label{11}
\mu_i^S(C_S, T, J_S) = \mu_i^{*S}  + \frac{\partial
\mu_i^{*S}}{\partial c_0}(C_{S} - c_0) +
 \frac{\partial \mu_i^{*S}}{\partial T^*}(T - T^*)\,,
\end{equation}
where
\begin{equation}\label{12}
   \mu_i^{*LS}  =  \mu_i^{LS}(C_{LS} = c_0, T = T^*,  J_{LS} =
0) = \mu_{leq,\,i}^{LS} (c_0, T^*)
\end{equation}
Further let us  consider  a dilute solution for which Raoult's and
Henry's laws are valid.  In this case local equilibrium chemical
potentials can be represented as
\begin{eqnarray}
  &&\mu_{leq,\,A}^{LS} (C, T) = \mu_{0A}^{LS}(T) + RT\ln(1 - C)\label{13}\\
   && \mu_{leq,\,B}^{LS} (C, T) = \mu_{0B}^{LS}(T) + RT\ln C\,,\label{14}
\end{eqnarray}
where $\mu_{0A(B)}^{L(S)}(T)$ is chemical potential of pure
component $A(B)$ at the temperature $T$ in the liquid (solid) state.
Calculating the derivatives of  (\ref{13}) - (\ref{14}) and
substituting them in (\ref{10}) - (\ref{11}), one obtains (at $J_S =
0$)
\begin{eqnarray}
\mu^{LS}_A =  \mu^{*LS}_A -
\frac{RT^*}{1 - c_0}(C_{LS} - c_0)\phantom{aaaaaaaaaaaaaaaa}&&  \nonumber \\
+ \; [-S^{*LS}_{0A} + R\ln (1 - c_0)](T - T^*) +
\alpha_A\frac{RT^*}{\rho V_D}J_{LS}&&\label{15}\\
 \mu^{LS}_B = \mu^{*LS}_B -
\frac{RT^*}{c_0}(C_{LS} - c_0) \phantom{aaaaaaaaaaaaaaaa}&&  \nonumber \\
+ [-S^{*LS}_{0B} + R\ln c_0](T - T^*) + \alpha_B\frac{RT^*}{\rho
V_D}J_{LS}\,,&&\label{16}
\end{eqnarray}
where $S^{*LS}_{0i} = - \partial\mu^*_{0i}/\partial T^*$ is entropy
pure component $i$ in local equilibrium at temperature $T^*$. It
follows from  (\ref{15})-(\ref{16}) that the local nonequilibrium
chemical potentials changes through the interface are
\begin{eqnarray}
\Delta\mu_A =  \Delta\mu_A^* + L_A^*(T - T^*)/T^*
\phantom{aaaaaaaaaaaaaaaa}&&  \nonumber \\
+ RT^*\frac{C_L - C_S}{1 - c_0} - \alpha_A\frac{RT^*}{\rho V_D}J_L \label{17}\\
\Delta\mu_B =  \Delta\mu_B^* + L_B^*(T - T^*)/T^*
\phantom{aaaaaaaaaaaaaaaa}&&\nonumber \\
 + RT^*\frac{C_S - C_L}{c_0} - \alpha_B\frac{RT^*}{\rho V_D}J_L &&\,,\label{18}
\end{eqnarray}
where $L_i^* = - T^*(S_{0i}^{*S} - S_{0i}^{*L})$ is the latent heat
solidification of the pure component $i$ at the temperature $T^*$
($L_i^*
> 0$), $\Delta\mu_i^* = \mu_i^{*S} - \mu_i^{*L} = \mu_{0i}^{*S} - \mu_{0i}^{*L} = \Delta\mu_{0i}^*$ and $\mu^{*LS}_{0i} = \mu^{LS}_{0i}(T^*)$.

\section{The velocity-temperature response function}
\subsection{  The interface movement without solute drag}
\label{3.1}
 Taking into account Eqs. (\ref{17})-(\ref{18}), one writes
down the equality (\ref{7}) for model without solute drag in the
form
\begin{eqnarray}
&& (1 - C_S) \Big[\Delta\mu_{0A}^* +  L_A^*(T - T^*)/T^* \nonumber\\
&&+ RT^*\frac{(C_L - C_S)}{1 - c_0} - \alpha_A\frac{RT^*}{\rho V_D}J_L\Big] +\nonumber\\
&& + C_S\Big[\Delta\mu_{0B}^* + L_B^*(T - T^*)/T^*   + \frac{RT^*}{c_0}(C_S - C_L) \nonumber\\
&&- \alpha_B\frac{RT^*}{\rho V_D}J_L\Big]
 =   RT\ln(1 - V/V_0) \label{19}
\end{eqnarray}
It follows from   (\ref{19}) that at $V = V_D$, $ C_L = C_S = c_0$,
$ T = T^*$ and $J_L = 0$ there is the equality
\begin{equation}\label{20}
    (1 - c_0)\Delta\mu_{0A}(T^*) + c_0\Delta\mu_{0B}(T^*) =
    RT^*\ln(1 - V_D/V_0)\,,
\end{equation}
which can be considered as an equation for determining temperature
$T^*$ of the interface  moving at the speed $V_D$ at the initial
concentration of the solute in the melt $c_0$. It can be shown that
for a diluted melt and a small deviation $ T^* - T_A$, where $ T_A $
is the melting temperature of the major component, the solution of
this equation coincides with the interface temperature $T^*$ found
in the works \cite{G02,S15} (see Appendix).

Taking into account the Eq. (\ref{20}) and the boundary condition
$J_L = (C_L - C_S)\rho V$, one can rewrite the Eq. (\ref{19}) in the
form
\begin{eqnarray}
& (c_0 -  C_S)(\Delta\mu_{0A}^* - \Delta\mu_{0B}^*) + L^*(C_S)(T - T^*)/T^* & \nonumber \\
&{\displaystyle +  \frac{RT^*(C_L - C_S)(c_0 - C_S)}{c_0(1 - c_0)}} & \nonumber \\
&- \alpha (C_S)RT^*(C_L - C_S)V/V_D& \nonumber \\
 & = RT\ln(1 - V/V_0) - RT^*\ln(1 - V_D/V_0) ,&
  \label{21}
\end{eqnarray}
where the notations are introduced
\begin{eqnarray}
   && L^*(C_S) = (1 - C_S)L^*_A + C_SL^*_B\nonumber\\
   &&  \alpha (C_S) = (1 - C_S)\alpha_A + C_S\alpha_B\,.\nonumber
\end{eqnarray}
In the absence of diffusion in the solid  one can put $C_S = c_0$
and write Eq. (\ref{21}) in the linear approximation in $V - V_D$
and $T - T^*$ as
\begin{equation}\label{22}
    \frac{T - T^*}{T^*} = \frac{\alpha (1 - k)C_L + (V_D - V)(V_0 - V_D)^{-1}}{L^*/RT^* -\ln(1 -
    V_D/V_0)}\,,
\end{equation}
where  $k = C_S/C_L = c_0/C_L$ is the nonequilibrium solute
partition coefficient and  $L^*$, $\alpha$  are taken at $C_S =
c_0$. In the derivation of Eq. (\ref{22}) it has been taken into
account that $1 - k \sim V_D - V$ at $V < V_D$. The equality
(\ref{22}) is the sought response function  representing kinetic
phase diagram in the case of  the interface movement with a speed
close to $V_D$. At $V_D/V_0 \ll 1$ one can simplify the expression
(\ref{22})  and get the nonequilibrium liquidus  equation in the
form
\begin{equation}\label{23}
    T = T^* + \alpha\frac{ RT^{*2}}{L^*}(1 - k)C_L + \frac{ RT^{*2}}{L^*}\frac{V_D - V}{V_0}
\end{equation}
In the linear approximation in $ V - V_D$ the solidus and liquidus
lines coincide since in this case $ (1 - k) C_L = (1 - k) C_S
/k\simeq (1 - K) C_S $.

In the region of the speeds $V$ close to $V_D$ the partition
coefficient $k(V)$ can also be  expanded into a series in powers of
$V_D - V$. Since for the interface the only dimensionless quantity
including $V_D$ is the ratio $V_{DI}/D_D$, where $V_{DI}$ is the
interfacial diffusion speed, then for dimensional reasons the
expansion of $k(V)$ in this region can be written as
\begin{equation}\label{24}
k(V)= \left\{
\begin{array}
{lll} 1 - f(V_{DI}/V_D)(1 - V/V_D) + \cdots \,,  &V <
V_D \\
1,  &V \geq V_D
\end{array}\right .\,,
\end{equation}
where  $f$ is some dimensionless function, in the simplest case one
takes $f(V_{DI}/V_D) = f_0V_{DI}/V_D$ and $f_0
> 0$ is dimensionless coefficient of the order of one. It should be
noted that Eq.~(\ref{24}) represents the expansion of $k(V)$ near
$V_D$ in the most general form regardless of the details of the
interface kinetics. The different representations of the
velocity-dependent partition coefficient for some models of
interface kinetics, taking into account complete solute trapping  at
$V \geq V_D$, can be found in Refs. \cite{G07, S12}. For example,
for the partition coefficient from \cite{G07} $f(V_{DI}/V_D) = 2(1 -
k_e)(1 - c_0)V_{DI}/V_D$, where $k_e$ is the equilibrium partition
coefficient.

Substitution of (\ref{24}) into (\ref{23}) gives in the linear
approximation in $V - V_D$
\begin{eqnarray}\label{25}
\Delta T = (V_D/V_0)\times\phantom{aaaaaaaaaaaaaaaaaaa}\nonumber&&\\
\times\left\{
\begin{array}
{lll} (1 + \alpha c_0f_0V_{DI}V_0/V_D^2)(1 - V/V_D),  &V <
V_D \\
 1 - V/V_D,   &V \geq V_D
\end{array},\right .&&\nonumber\\
\end{eqnarray}
where $\Delta T = (T - T^*)L^*/RT^{*2}$. As it can be seen from
(\ref{25}), at positive $\alpha$, $\Delta T > 0$ for $V/V_D < 1$ and
$T$ decreases with the growth of $V$ changing the slope at the point
$V_D$ (the behavior of $T$ is qualitatively  given by  the curve 1
in Fig.\ref{fig1}).

\subsection{Solute drag effect}
\label{3.2}

\begin{figure}[t]\centering
    \includegraphics[width= 0.45\textwidth]{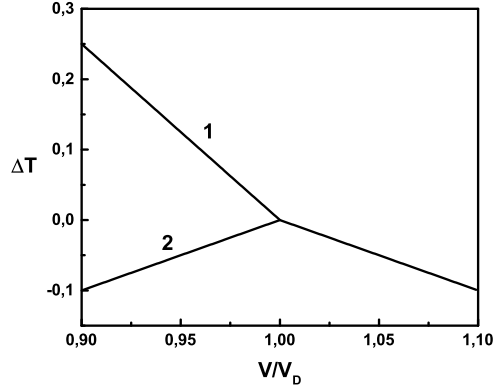}
\caption{  The qualitative behavior of $ \Delta T $ depending on $ V
/ V_D $; the curve 1 corresponds to the Eq.(\ref{25}) (model without
solute drag); the curve  2 corresponds to the Eq.(\ref{30}) (model
with solute drag) at $|\Delta\mu^*_{0A}|V_0/10RT^*V_D > 1$
\label{fig1}}
      \end{figure}

In the  model with  drag effect one has
\begin{equation}\label{25_1}
 C_L\Delta \mu_B + (1 - C_L\Delta\mu_A) = RT\ln (1 -
 V/V_0)\,.
\end{equation}
Substituting now  relations (\ref{17}) and (\ref{18}) in
Eq.(\ref{25_1}) and performing transformations similar to those used
in the derivation of  Eq.(\ref{22}), one obtains in the linear
approximation in $V - V_D$
\begin{eqnarray}\label{26}
    \frac{T - T^*}{T^*} = \phantom{aaaaaaaaaaaaaaaaaaa}&&\nonumber \\
    \frac{(\alpha + \Delta\mu^*_0/RT^* )(1 - k)C_L + (V_D - V)(V_0 - V_D)^{-1}}{L^*/RT^* -\ln(1 -
    V_D/V_0)}\,,&&
\end{eqnarray}
where $\Delta\mu^*_0 = \Delta\mu_{0A}^* - \Delta\mu_{0B}^*$.  At
$V_D/V_0 \ll 1$  Eq. (\ref{26}) yields the equation of  the
nonequilibrium liquidus in the form
\begin{equation}\label{27}
    T = T^* + \frac{ RT^{*2}}{L^*}(\alpha + \Delta\mu^*_0/RT^*)(1 - k)C_L + \frac{ RT^{*2}}{L^*}\frac{V_D -
    V}{V_0}\,.
\end{equation}
In the linear approximation in $V - V_D$ the solidus equation is
also given by the Eq. (\ref{27}) (see the note after Eq.
(\ref{23})). At $V \geq V_D$ the Eq.(\ref{27}) and Eq.(\ref{23})
coincide. To analyze the Eq.(\ref{27}) for $V < V_D $ we consider
the Eq.(\ref{20}) from which it follows that

\begin{equation}\label{28}
   \frac{\Delta\mu^*_0}{RT^*} =  \frac{\Delta\mu_{0A}^*}{c_0RT^*} +
   \frac{V_D}{c_0V_0}\,.
\end{equation}
Taking into account the Eq.(\ref{28}) one can rewrite the
Eq.(\ref{27}) as
\begin{equation}\label{29}
 \Delta T = \Big(\alpha c_0 + \frac{\Delta\mu^*_{0A}}{RT^*} +
 V_D/V_0\Big)\frac{1 - k}{k} + \frac{V_D - V}{V_0}\,.
\end{equation}
Bearing in mind the diluted melt ($ c_0\rightarrow 0 $) and using
(\ref {24}), we obtain from (\ref {29}) in the linear approximation
in $V - V_D$ (by neglecting the second-order term $c_0(V - V_D)$)
\begin{equation}\label{30}
 \Delta T = \Big[f_0\frac{V_{DI}}{V_D}\Big(\frac{\Delta\mu^*_{0A}}{RT^*} + \frac{V_D}{V_0}\Big)
 + \frac{V_D}{V_0}\Big](1  - V/V_D)\,.
\end{equation}
It is seen from (\ref {30}) that for $f_0\approx 1$ the inequality
$\Delta T >0 $ holds\footnote{At $T = T^* < T_A$, where $T_A$ is the
melting temperature  of major component, the solid phase is more
stable and $\Delta\mu^*_{0A} = \mu^{*S}_{0A} - \mu^{*L}_{0A} < 0$}
 if
\begin{equation}\label{31}
\frac{|\Delta\mu^*_{0A}|}{RT^*}\frac{V_0/V_D}{1 + V_D/V_{DI}} < 1\,.
\end{equation}
Under the condition $V_0/V_D \gg 1$, the expression on the left side
of inequality (\ref {31}) can be large enough (for example, for
Si-9\, at.\% As $V_D/V_{DI} \sim 1$ \cite{HGH07}). Therefore, it is
obvious that the condition (\ref {31}) will not be automatically
satisfied for all systems. This means that when the speed of the
front increases towards the critical value of $V_D$ in some cases
solute drag effect can be accompanied by an increase in the
interface temperature, $\Delta T = T-T^* < 0$ (curve 2 in
Fig.\ref{fig1}). It is this behavior that was found in the numerical
simulation of the rapid solidification of the Si-9 at.$\%$ As system
\cite{LZ10}.

\section{Conclusion}
The most important consequence of the absence of local equilibrium
in the bulk of the liquid phase in the processes of rapid
solidification is transition to complete solute trapping at the
finite velocity of the solidification front $V = V_D$.  This
circumstance makes it possible to analyze analytically the details
of the solidification process in the region of high velocities close
to $V_D$.  For the indicated speed range the analytical expression
for the temperature response function representing kinetic phase
diagram and taking into account solute drag effect has been derived.
 In contrast to previous works, in the given approach the equilibrium properties of the alloy based
 on the equilibrium phase diagram of the system have not been used. At high speeds of the solidification front, the
 interface temperature may be quite low, so that in the  equilibrium state at this  temperature  the alloy can only exist in the form of a solid solution  and the use of equilibrium liquidus and solidus  loses  its meaning.

In the case of a dilute melt, as it follows from the analysis of the
response function (\ref{30}), solute drag can be accompanied by an
increase of the  temperature of the interface when its velocity
approaches to the critical value of $V_D$. Numerical simulation of
high-speed solidification of Si-9 at.$\%$ As alloy   shows also the
existence of a local temperature maximum in the region of large $V$
\cite{LZ10}. Growth of the interface temperature as  $V$  approaches
to $V_D$   is perhaps some sign of  drag effect. However detailed
research on this issue requires studying the solidification process
on a large number of different binary systems.

\section{Appendix}
\label{app}
 The solution of the
equation (\ref{20}) for the interface temperature moving with
velocity $V = V_D$ can be relatively easily found  for the case of a
dilute melt, $c_0 \rightarrow 0$, and a small deviation $T^* - T_A$.
At $V_D/V_0 \rightarrow 0$ instead of Eq.(\ref{20}) one has
\begin{equation}\label{A1}
    (1 - c_0)\Delta\mu_{0A}(T^*) + c_0\Delta\mu_{0B}(T^*) =
   - RT^* V_D/V_0\,.
\end{equation}
 Let's present the solution of the last equation as
 \begin{equation}\label{A2}
   T^* = T_A +  a\,c_0 + b\,V_D/V_0\,,
\end{equation}
where $a$ and $b$  are some coefficients to be determined. Taking
into account that $\Delta\mu_{0A}(T_A) = \mu_{0A}^S(T_A) -
\mu_{0A}^L(T_A) = 0$, substitution of (\ref{A2}) into (\ref{A1})
gives in the linear approximation in $c_0$ and $V_D/V_0$
\begin{eqnarray}\label{A3}
    &&\Bigg[ a\,\bigg(\frac{d \Delta\mu_{0A}}{dT}\bigg)_{T_A} + \Delta\mu_{0B}(T_A )
    \Bigg]c_0\nonumber\\
    &&\phantom{aaaaaa} + \Bigg[b\,\bigg(\frac{d \Delta\mu_{0A}}{dT}\bigg)_{T_A} +
    RT_A\Bigg]\frac{V_D}{V_0} = 0\,.
\end{eqnarray}
Provided that $c_0$ and $V_D/V_0$ are independent quantities, it
follows from Eq.(\ref{A3})
\begin{equation}\label{A4}
     a = - \frac{\Delta\mu_{0B}(T_A )}{\left(d\Delta\mu_{0A}/dT\right)_{T_A}}\,, \hspace{1cm} b = -
     \frac{RT_A}{\left(d\Delta\mu_{0A}/dT\right)_{T_A}}\,.
\end{equation}
Let's find the derivative from  (\ref{A4})
\begin{equation}\label{A41}
\frac{d \Delta\mu_{0A}}{dT} = \frac{d \Delta\mu_{0A}^S}{dT} -
\frac{d \Delta\mu_{0A}^L}{dT} = S_{0A}^L(T) - S_{0A}^S(T)\,,
\end{equation}
 It follows from this equation that
\begin{equation}\label{A5}
    \left(\frac{d \Delta\mu_{0A}}{dT}\right)_{T_A} =
    \frac{L_A}{T_A}\,,
\end{equation}
where  $L_A = T_A (S_{0A}^L - S_{0A}^S)$ is the latent heat
solidification of   the pure component $A$, and
\begin{equation}\label{A6}
    b = - \frac{RT_A^2}{L_A}
\end{equation}

To find $\Delta\mu_{0B}(T_A )$ consider the temperature $T$, close
to $T_A$, at which the liquid and solid phases of the binary system
are in equilibrium with each other with concentrations equal,
respectively to $C_L^e$  and $C_S^e$, so that $\mu^L_{leq,B}(T) =
\mu^S_{leq,B}(T)$. Then using equality (\ref{14}), one gets
\begin{equation}\label{A61}
   \ln k_e = \ln\frac{C_S^e}{C_L^e} = \frac{\mu_{0B}^S(T) -
   \mu_{0B}^L(T)}{RT} = - \frac{\Delta\mu_{0B}(T )}{RT}\,,
\end{equation}
where $k_e = C_S^e/C_L^e$ is the equilibrium partition coefficient,
and for $T \rightarrow T_A$,
\begin{equation}\label{A7}
\Delta\mu_{0B}(T_A) = - RT_A\ln k_e\,.
\end{equation}
Substituting (\ref{A7}) into $a$ from (\ref{A4}) and taking into
account (\ref{A5}), one obtains
\begin{equation}\label{A8}
    a = - \frac{RT_A^2\ln k_e}{L_A}\,.
\end{equation}
Using the well-known equality $RT_A^2/L_A = m_e/(k_e - 1)$, where
$m_e$ is the equilibrium slope of the liquidus line, and relations
(\ref{A6}) and (\ref{A8}), one can  write the interface temperature
finally in the form
\begin{equation}\label{A9}
 T^* = T_A +  \frac{c_0\, m_e}{k_e - 1}\ln k_e -  \frac{ m_e}{k_e -
 1}\frac{V_D}{V_0}\,.
\end{equation}
Eq.(\ref{A9}) coincides with  the expression  for the temperature of
the interface moving with the speed $V_D$ found in works
\cite{G02,S15} for a dilute melt.




\end{document}